\documentclass[aps,amssymb,amsmath,prb,reprint,noshowpacs,]{revtex4-1}
\usepackage{times}
\usepackage{amsmath}
\usepackage{graphicx}
\usepackage{epsfig}
\usepackage{bm}

\usepackage{scalefnt}   

\begin{document}

\title{
Experimental evidence for large dynamic effects on the plasmon
dispersion of subwavelength metal nanoparticle waveguides }

\author{A. Femius Koenderink}
\email{f.koenderink@amolf.nl}
\author{Ren\'e de Waele}
\author{Jord C. Prangsma}
\author{Albert Polman}
\affiliation{Center for Nanophotonics, FOM Institute for Atomic and
Molecular Physics,  Kruislaan 407, 1098 SJ Amsterdam, The
Netherlands}

\date{Published as Phys. Rev. B \textbf{76}, 201403(R) (2007). [Received July 18, 2007, published November 13, 2007)]}

\begin{abstract}

We present angle  and frequency resolved  optical extinction
measurements to determine the dispersion relation of plasmon modes
on Ag and Au nanoparticle chains with pitches down to 75~nm. The
large splitting between transverse and longitudinal modes and the
band curvature are inconsistent with reported electrostatic
near-field models, and confirm that far-field retarded interactions
are important, even for $\lambda/5$-sized structures. The data imply
that lower propagation losses, larger signal bandwidth and larger
maximum group velocity then expected can be achieved for wave
vectors below the light line. We conclude that for the design of
optical nanocircuits coherent far-field couplings across the entire
circuit need to be considered, even at subwavelength feature sizes.
\end{abstract}


 \maketitle

A fundamental limit to the realization of sub-wavelength
(sub-$\lambda$) optical devices is that the interaction strength of
dielectric objects with light vanishes as the objects gets
smaller.\cite{jackson,huffman} Plasmonics may allow to overcome this
inherent limitation of dielectrics by packing the large
polarizability of free electron resonances  into a small physical
volume.\cite{huffman,schatzanisotrop,plasmonbook} In this framework
plasmon particle arrays have been proposed as an ideal platform that
combines the ease of controlled nanofabrication with the prospects
of creating, e.g., ultra-small antennas to efficiently harvest,
enhance and emit optical power,\cite{hernandez,waele} as well as a
toolkit for nanophotonic circuits.\cite{enghetaprl06} Thus plasmon
chains may act as sub-$\lambda$ width waveguides, waveguide bends,
signal splitters, and filters.\cite{quinten,brongersma,maier,park}
As a parallel development, sub-$\lambda$ arrays of scatterers with
magnetic rather than electric resonances have recently gained
tremendous interest for developing optical
metamaterials.\cite{metalist}

Pioneering experiments have focused on  qualitative understanding of
the resonance splitting and mode structure in plasmon particle
clusters and arrays.\cite{maier} The observed polarization-dependent
resonances in linear (1D) particle chains, for instance, correlated
well with trends anticipated from a simple near-field
quasi-electrostatic model for the chain dispersion
relation.\cite{brongersma,park} This   `quasistatic' model, which is
valid on deep sub-$\lambda$ length scales, also formed the basis for
forecasting the functionality of more complex structures, such as
plasmon chain splitters and
multiplexers.\cite{brongersma,enghetaprl06,multiplexer} Very
recently, however, several groups have developed electrodynamic
models that predict large quantitative and qualitative deviations
from the quasistatic
insights.\cite{zhaojpcb_MP,simovskicitrin,weber,koenderink06,alu} If
these deviations indeed occur, a redevelopment of general design
rules for complex nanophotonic circuits is required to include
electrodynamic effects even on small length scales.

Sofar, quantitative experiments to discriminate between quasistatic
and electrodynamic predictions  at sub-$\lambda$ spacings have not
been reported. In this Letter we present   angle  and frequency
resolved optical extinction experiments on many arrays of Ag and Au
nanoparticles at various particle sizes and sub-$\lambda$ pitches
down to $\lambda/5$. Sub-$\lambda$ plasmon arrays are an interesting
system to test for dynamic effects. Recently, large modifications
were predicted for the  dispersion of modes in 1D plasmon particle
chains and, equivalently, in 1D magnetic split ring resonator
arrays.\cite{magnetic1D} For instance, the deviations include a much
larger $k=0$ splitting between the longitudinal and transverse
modes, and a polariton splitting of the transverse dispersion branch
at the crossing with the vacuum dispersion relation (light
line).\cite{simovskicitrin,weber,koenderink06,alu} For sub-$\lambda$
waveguiding these models further imply a large increase in group
velocity and decrease in propagation loss compared to electrostatic
predictions.\cite{koenderink06} The dispersion relations measured in
our experiment agree well with the recent electrodynamic models, and
deviate strongly from quasistatic
 predictions down to pitches as small as $\lambda/5$.
As a consequence we anticipate that design proposals for
sub-$\lambda$ nano-optical circuits at currently realistic sizes
($\sim 50$~nm pitch) can not be based on quasistatic
analysis,\cite{enghetaprl06} because coherences and coupling across
the full structure will dominate the optical performance.

\begin{figure}[b]
\centerline{\includegraphics[width=0.81\columnwidth]{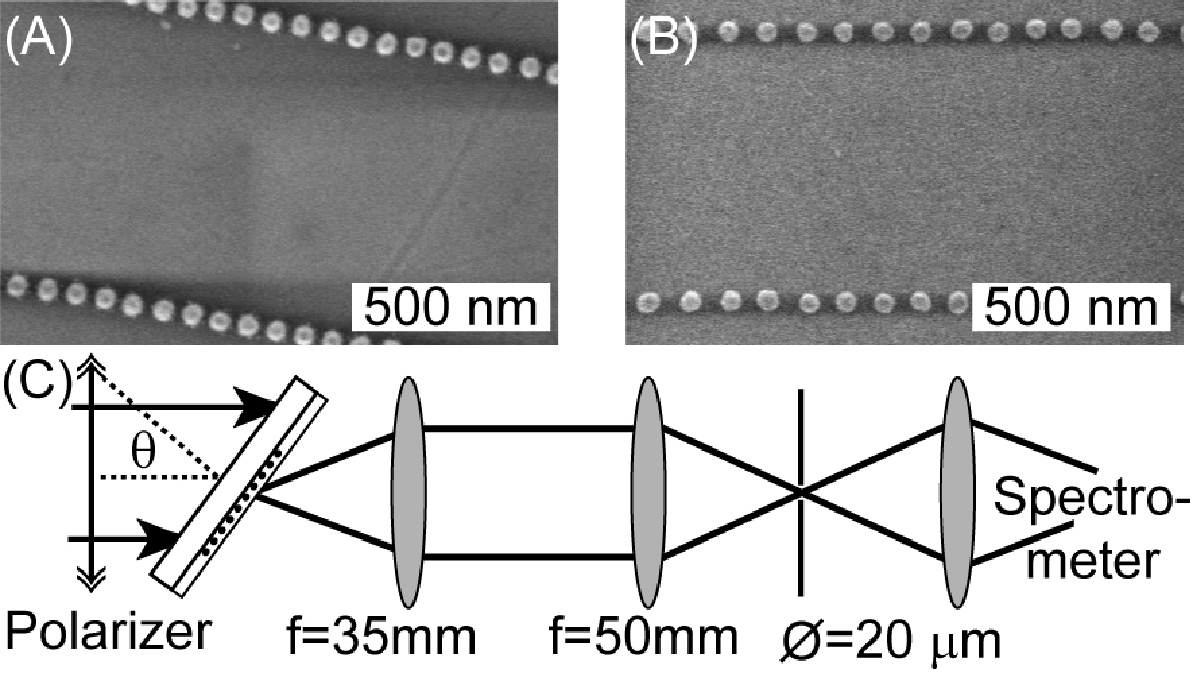}}
\caption{(A, B) Scanning electron micrographs of particle arrays on
glass with $r=25$~nm, $d=75$~nm (A, Au) and $r=25$, $d=100$ nm (B,
Ag). (C) Overview of the angle-dependent transmission setup. A
collimated beam illuminates a large sample area. The sample angle is
varied. Light is collected from a small spot on the sample
(collection NA$=0.2$). \label{fig1:sem}}
\end{figure}

We determine the nanoparticle chain dispersion relation above the
light line by far-field  extinction measurements on Ag and Au
nanoparticle arrays prepared on glass using electron-beam
lithography.  After
physical vapor deposition of Ag or Au and resist liftoff, we 
obtained linear arrays of particles of 50 nm height, at pitches of
$d=75, 100, 120$ and $150$~nm and with radii varied between $r=25$
and $55$~nm, as determined by scanning electron microscopy (SEM,
cf.~Fig.~\ref{fig1:sem}(A,B)). The estimated error in determining
$r$ is 2~nm. For each pitch the particle radii are below $r/d=0.37$.
We have fabricated square fields containing  parallel particle
chains (chain length $60~\mu$m), with randomly varying inter-chain
spacing (minimum 700~nm, mean 1 $\mu$m). The large inter-chain
spacing ensures that coupling between
  chains is negligible, while the random variation suppresses
grating effects that occur  for periodic arrangements. Finally, we
spin-coat the samples with a 100~nm layer of PMMA to ensure that the
particle chains are embedded in a homogeneous dielectric
environment. To determine the dispersion relation we use a
wavelength-resolved transmission setup (see Fig.~\ref{fig1:sem}(C))
in which the sample is mounted on a rotation stage (axis
perpendicular to the chains, along the PMMA/glass interface) that
gives access to incident angles from $\theta=-60^\circ$ to
$+60^\circ$. The chains are illuminated by a   collimated
white-light beam (divergence $\sim 5^\circ$) from a fiber-coupled
incandescent source, which illuminates a large (mm-size) area on the
sample. Using a pinhole on the transmission side, only the
transmitted intensity from a $\sim 15~\mu$m spot
(associated $\Delta k/k \sim  0.09$ (i.e., 5$^\circ$)) 
 on the sample
is collected by a cooled Si-CCD coupled spectrometer. The
transmission is obtained by normalizing the transmitted intensity to
that recorded from an unpatterned substrate at the same angle. Using
a broadband polarizer, we  select the incoming polarization to be
either transverse
to the chains, or longitudinal (p-polarization). For nonzero
$\theta$, the p-polarization also acquires a field-component
transverse to the chain.

\begin{figure}[tb]
\centerline{\includegraphics[width=0.9\columnwidth]{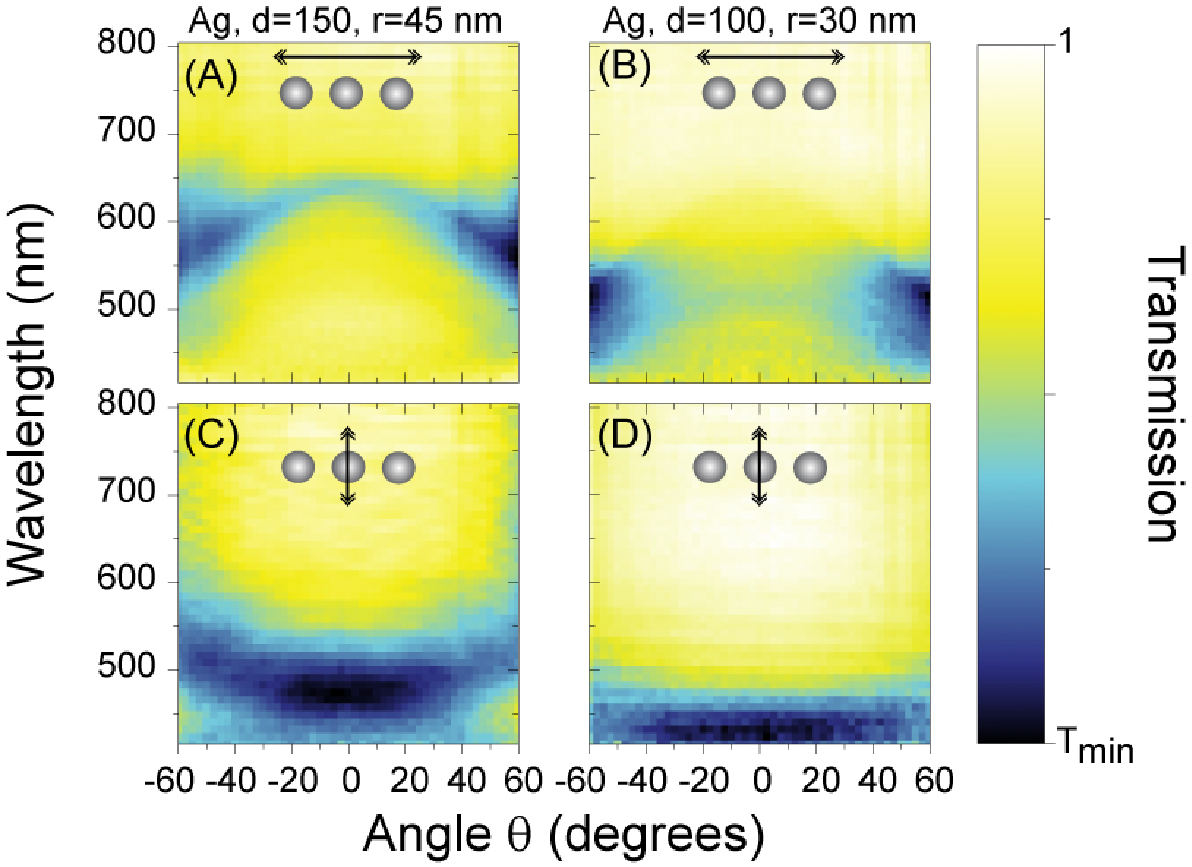}}
\caption{(Color online) Transmission   as a function of angle  and
wavelength for longitudinal  (A,B) and transverse polarization (C,D)
for Ag arrays with $r=45, d=150$~nm (A,C) and $r=30,d=100$~nm (B,D).
The color scale runs from $T_{\rm min}=0.65$ in (A,C) and $T_{\rm
min}=0.80$ in (B,D).\label{fig2:spectra}}
\end{figure}
Figure~\ref{fig2:spectra} shows transmission spectra for the full
angular range for Ag   chains of pitch 150~nm, and radius 45 nm
(Fig~\ref{fig2:spectra}(A,C)), and for a smaller pitch of 100~nm and
radius 30~nm (Fig~\ref{fig2:spectra}(B,D)). At normal incidence
($\theta=0^\circ$), a band of extinction around $\lambda=500$~nm for
$d=150$~nm (470~nm for $d=100$~nm) is observed for transverse
polarization, and at $\lambda=625$~nm ($550$ nm for $d=100$~nm) for
longitudinal polarization. The redshift of modes for larger
particles is consistent with the well-known single-particle
resonance shift with particle size.\cite{huffman,carron} The
occurrence of two bands shifted to either side of the single
particle resonance is consistent with earlier reports on the
 $\theta=0^\circ$ extinction of nanoparticle chains.\cite{maier}
Qualitatively this splitting corresponds to the excitation of
collective modes in the chain of dipole scatterers: the transverse
mode is blue shifted due to the antiparallel orientation of each
dipole with the field of its neighbors, while the longitudinal mode
is red shifted as each dipole is aligned with the field of its
neighbors. For increasing angle of incidence, the two branches have
opposite curvature, both shifting towards the single particle
resonance.

\begin{figure}[tb]
\centerline{\includegraphics[width=0.85\columnwidth]{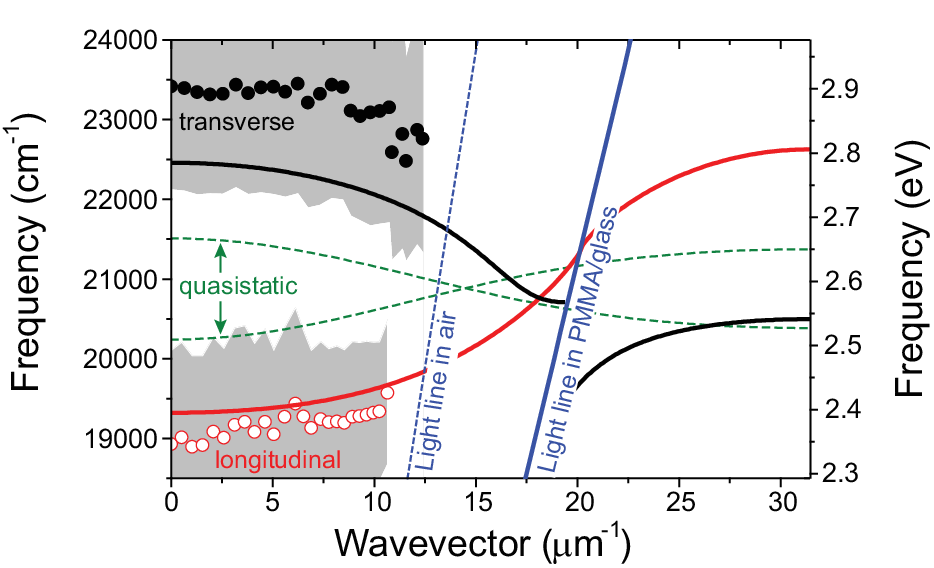}}
\caption{(Color online) Symbols: frequency (in cm$^-1$ (left axis)
or eV (right axis)) of minimum transmission versus wave vector for
transverse and longitudinal polarization in
Fig.~\ref{fig2:spectra}(B,D), i.e. the $d=100~$nm, $r=30$~nm Ag
particles. Shaded areas: 1/e bandwidth of the transmission minima.
Straight lines: light lines in air and in the embedding medium.
Thick (thin) curves: dynamic (static) prediction.
\label{fig3:kparplot}}
\end{figure}

The spectral dependencies of the extinction branches evident in
Fig.~\ref{fig2:spectra} are qualitatively consistent with both the
quasistatic model and full dynamic calculations for the dispersion
relation of plasmon chain excitations. In order to quantitatively
compare the data with the two models, we determine the transmission
minima from a Gaussian fit to the transmission spectrum (plotted in
the frequency domain) for each angle. In figure~\ref{fig3:kparplot},
we plot the resulting center frequencies taken from
Fig.~\ref{fig2:spectra}(B,D) as a function of
$|\mathbf{k}_{||}|=2\pi/\lambda \sin(|\theta|)$, i.e. the wave
vector component of the incident beam along the chain. The
measurements reach up to $k_{||}=0.86 \omega/c$, or up to 60\% of
the light line in the medium embedding the particles. To appreciate
the large width of the extinction resonances, the bandwidth at $1/e$
height of the fitted Gaussians is shaded in the diagram. The typical
$1/e$ full width is 2500 cm$^{-1}$  ($\sim 60$ nm). First, we
compare to the generic quasistatic point-dipole prediction for the
nanoparticle chain dispersion relation:\cite{maier,brongersma,park}
\begin{equation}\omega^2 =\omega_0^2[1 + \left(\frac{r}{d}\right)^3
\sum_{j=1}^\infty\kappa_{T,L} \frac{\cos(jkd)}{j^3} ].
\label{eq:quasi}\end{equation} Here $\omega_0$ is the
single-particle resonance frequency, and  $\kappa_T=2$ for
transverse, and $\kappa_L=-4$ for longitudinal
modes.\cite{modifieddrudeQ}  We assume spherical particles with
$r/d$ taken from SEM data to obtain the quasistatic prediction  in
Fig.~\ref{fig3:kparplot} (dotted curves). There are two striking
discrepancies between the data and the quasistatic model. First, the
splitting at normal incidence ($k_{||}=0$) between the two branches
is a factor two to three larger in the data than in the quasistatic
model. The quasistatic model even falls outside the broad width of
the extinction peaks. Secondly, the quasistatic model predicts that
the transverse and longitudinal branch cross at $kd=0.46\pi$ {\rm
independent} of $r/d$ at a frequency equal to the single-particle
resonance. No sign of this crossing is observed in the data. It
seems surprising that such discrepancies between data and the
quasistatic model haven't been noted in earlier studies.\cite{maier}
These studies focused on $k=0$ only,   without investigating nonzero
scattering angles. On the basis of $k=0$ data only, one might assume
that a larger splitting is due to an error in $r/d$. For our data,
this would imply an unlikely 40\% error in estimating $r/d$.
However, even if one would scale $r/d$ to match the  $k_{||}=0$
splitting,  the quasistatic model would still not be consistent with
the full angle-dependent data set: the presence of the band crossing
at $kd=0.46\pi$ would impose a much larger curvature of both bands
than observed in our data.

The fact that the quasistatic model does not describe the data could
be due to several approximations: The quasistatic model ignores
dynamic effects,  multipole effects, the presence of the PMMA-air
interface, and particle anisotropy. Based on Ref.~\onlinecite{park}
we conclude that  multipole effects do not resolve the discrepancy
between the quasistatic model and our data: multipole effects do not
alter the quasistatic dispersion relation for $r/d < 0.35$, a
criterion satisfied in our experiment ($r/d=0.3\pm 0.02$). For
larger $r/d$ multipole effects in fact reduce the quasistatic $k=0$
splitting, and shift the band crossing to even smaller wave
vectors,\cite{park} inconsistent with the absence of a crossing in
our data. To study the effect of the dielectric interface, we have
analyzed the quasistatic model for dipoles near a dielectric
interface using image dipole theory.\cite{jackson} Within this model
the image dipoles have a weaker dipole moment by a factor
$(\epsilon_m-1)/(\epsilon_m+1)$ (with $\epsilon_m$ the embedding
dielectric constant) and are located at a distance of 150 nm, i.e.,
further than the array pitch. As a consequence the estimated
increase in splitting is less than 15\%, i.e., much less than the 2
to 3-fold enhancement in the experiment. Regarding anisotropy, the
particles for which data is reported in Fig.~\ref{fig3:kparplot} are
mildly oblate (height 50~nm and diameter 60 nm). At
$\theta=0^\circ$, both polarizations used in the experiment are
along equivalent axes of the single-particle polarizability tensor,
so that an enhanced splitting is not due to excitation of distinct
single-particle resonances. In addition, we  calculated the
polarizability tensor (including dynamic depolarization
shifts\cite{schatzanisotrop,carron}). The resonance relative to that
of a $r=30$~nm spherical particle is only shifted by 4 nm, and the
on-resonance polarizability along the long axes is in fact 10\%
smaller. Within the quasistatic model, particle anisotropy is hence
 expected to reduce the splitting, rather than explaining the
observed enhancement.

Having excluded that modifications to the quasistatic model due to
multipole effects, interface corrections or particle anisotropy can
explain the dispersion relation observed in the experiment, we now
compare the  data to a full electrodynamic point-dipole
model\cite{koenderink06} that includes Ohmic damping, radiation
damping, and depolarization shifts in the single-particle
polarizability\cite{carron}, as well as all terms in the dipole
field. Within this model the dispersion relation has recently been
calculated perturbatively\cite{simovskicitrin}, and
self-consistently for finite\cite{weber} and infinite
arrays.\cite{koenderink06,alu} Using a dielectric model for silver
that is a modified Drude fit\cite{modifieddrude} to tabulated
data,\cite{palik} and taking $r$ and $d$ from SEM observations, we
plot the self-consistent infinite array model as solid curves in
Fig.~\ref{fig3:kparplot}. This model predicts a band splitting at
$k=0$ that is approximately twice the splitting in the quasistatic
model for the same geometrical parameters, as well as a band
curvature for wave vectors away from $k=0$ that is in much better
agreement with the experimental data. Keeping in mind that the
dynamic model has no adjustable parameters, the reasonable
correspondence with the data is a strong indication that retardation
effects and far-field coupling are indeed determining  the
dispersion relation for $d=100$~nm  silver nanoparticle chains. The
remaining deviation from the dynamic model may yet again be due to
the small particle anisotropy, the PMMA-air interface, or multipole
effects\cite{zhaojpcb_MP} the influence of which is much more
difficult to estimate quantitatively for electrodynamic rather than
quasistatic models.

\begin{figure}[tb]
\centerline{\includegraphics[width=0.85\columnwidth]{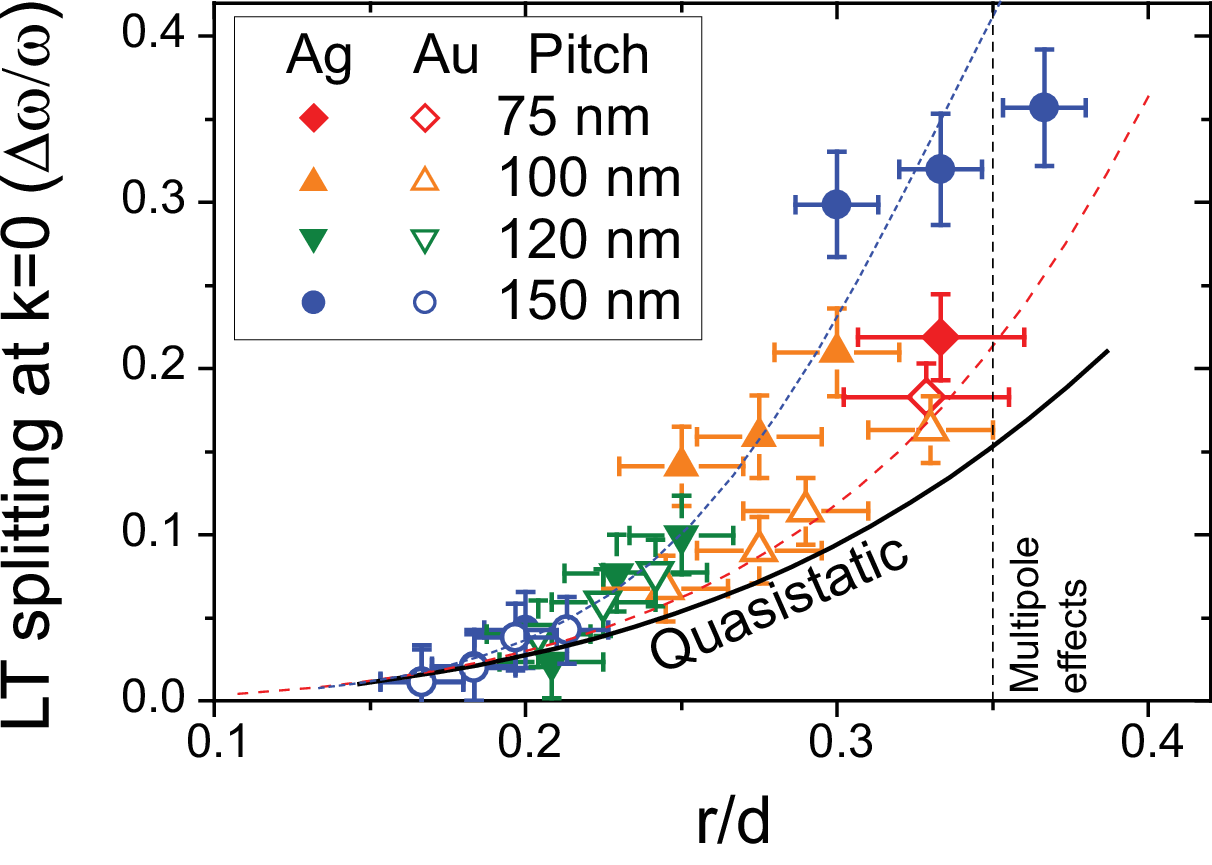}}
\caption{(Color online) Symbols: Relative splitting between
transverse and longitudinal branch at $k=0$ versus $r/d$ for several
pitches $d$ as indicated for Ag and Au arrays. Black curve:
quasistatic prediction for Ag arrays. Colored curves: dynamic
predictions for $d=75$~nm (Ag, dashed) and $d=150$~nm (Ag, dotted).
Multipole effects set in beyond the dashed vertical.\cite{park}
 \label{fig4:rdplot}}
\end{figure}
Based on the data in Figs.~\ref{fig2:spectra}, \ref{fig3:kparplot}
we conclude that the large splitting between the transverse and
longitudinal branch at normal incidence compared to the quasistatic
prediction is a good indicator for the relevance of far-field
effects. Figure~\ref{fig4:rdplot} shows the measured relative
splitting $\Delta\omega / \omega$ for many combinations of particle
radius $r$ and pitch $d$ plotted against $r/d$, both for Ag and Au
particle arrays. In the quasistatic limit, the splitting is simply
proportional to $(r/d)^3$ (see Eq. (\ref{eq:quasi})). In the dynamic
model, the splitting depends on $r$ and $d$ separately as is clear
from the dynamic predictions for the smallest ($d=75$~nm) and
largest ($d=150$~nm) pitch used in our experiments (dashed/dotted in
Fig.~\ref{fig4:rdplot}). Figure~\ref{fig4:rdplot} demonstrates that
for Ag particles the splitting is generally a factor $\sim 2$ larger
than quasistatic theory predicts,  in agreement with the dynamic
model. For Au particles, the difference is not as large, which may
explain why previous studies did not resolve deviations from
quasistatic theory.\cite{maier} Still, the observed splittings for
Au systematically exceed the quasistatic prediction. We attribute
this smaller effect for gold to the much lower albedo ($\leq 20$\%
for $r=25$~nm Au, compared to $\sim 80\%$ for Ag).\cite{huffman}
Since low-albedo particles radiate less strongly, far-field
corrections to the dispersion relation will be less important.

In conclusion, the experiment reported in this Letter shows that the
dispersion relation for plasmon modes on sub-$\lambda$ metal chains
is strongly modified by far-field interactions, even for pitches as
small as  $d=75~\mbox{nm}$ or $\sim \lambda/5$. This experiment thus
confirms recent models\cite{simovskicitrin,weber,koenderink06,alu}
that overturn the quasistatic view on plasmon
chains.\cite{brongersma,park} The next challenge is to address wave
vectors below the light line, for which propagation distances up to
5--10$~\mu$m at group velocities around $0.3c$ are
feasible.\cite{koenderink06} These propagation distances far exceed
initial estimates, which were based on the damping rate of single
particles.\cite{brongersma,maier} Below the light line, the damping
rate is strongly reduced since far-field destructive interference
suppresses all radiative loss. It will be a challenge to excite
these wave vectors selectively: local excitation, \emph{e.g.} at a
waveguide entrance, will excite wave vectors both below and above
the light line.  Recent simulations confirm that retardation effects
are very important for local excitation of finite sub-$\lambda$
plasmon chains, giving rise to a complex and strongly frequency
dependent response.\cite{hernandez} However, interpretation in terms
of superpositions of modes above and below the light line is
nontrivial, since depending on the excitation either the dispersion
relation for real $k$ and complex $\omega$, or  for complex $k$ with
real $\omega$ applies, as discussed in Ref.
\onlinecite{koenderink06}. As the complexity is increased to include
2D clusters, we expect far-field
 effects to become even stronger.\cite{zhaojpcb_MP}
Therefore, our work implies that  an essentially quasistatic
electric  circuit-design approach\cite{enghetaprl06} to
nanophotonics is only applicable to structures of total size below
$\sim 50~$nm, i.e., a scale at which fabrication of several coupled
components is extremely challenging. Sub-$\lambda$ photonic
structures at the current fabrication limit will always require a
fully electrodynamic optimization of the coherent coupling between
all building blocks: far-field interference is key to optimize the
functionality and reduce the loss.

We thank  L. Kuipers for loan of equipment and H.A. Atwater for
fruitful discussions. This work is part of the research program of
the ``Stichting voor Fundamenteel Onderzoek der Materie (FOM),''
which is financially supported by the ``Nederlandse Organisatie voor
Wetenschappelijk Onderzoek (NWO)''. It was also supported by
``NanoNed'', a nanotechnology program funded by the Dutch Ministry
of Economic Affairs.


\begin{thebibliography}{10}
\bibitem{modifieddrudeQ} For Drude models
with $\epsilon_b\neq 1, \gamma=0$,  $\kappa$ is reduced by
($3\epsilon_m/(\epsilon_b+2\epsilon_m)\approx 0.66$ for Ag  in
glass.
\bibitem{modifieddrude} We fit optical data in~\onlinecite{palik} using  $\epsilon_{\rm metal}=\epsilon_b
-\eta \omega_p^2/(\omega (\omega+i\gamma)$. For silver,
$\epsilon_b=5.45,\eta=0.73, \omega_p=1.72\cdot 10^{16}$~s$^{-1}$ and
$\gamma = 8.35\cdot 10^{13}$~s$^{-1}$.
\end{thebibliography}
\end{document}